# A new technique for laser cooling with superradiance

Galina Nemova[1,*] and Raman Kashyap[1,2]

*Department of Engineering Physics[1] and Department of Electrical Engineering[2],
École Polytechnique de Montréal, P.O. Box 6079, Station Centre-ville, Montréal, Canada*

[*]*Corresponding author:* galina.nemova@polymtl.ca

**Abstract:** We present a new theoretical scheme for laser cooling of rare earth doped solids with optical super-radiance (SR), which is the coherent, sharply directed spontaneous emission of photons by a system of laser excited rare earth ions in the solid state host (glass or crystal). We consider an $Yb^{3+}$ doped ZBLAN sample pumped at the wavelength 1015 *nm* with a rectangular pulsed source with a power of ~433W and duration of 10*ns*. The intensity of the SR is proportional to the *square* of the number of excited ions. This unique feature of SR permits a dramatic increase in the rate of the cooling process in comparison with the traditional laser cooling of the rare earth doped solids with anti-Stokes spontaneous incoherent radiation (fluorescence). This scheme overcomes the limitation of using only low phonon energy hosts for laser cooling.

## I. INTRODUCTION

At the present time laser cooling of solids is one of the most promising and significant problems of laser physics. The idea of cooling solid state matter with anti-Stokes fluorescence was first proposed by Pringsheim in 1929.[1] It has been shown that some materials emitted light at shorter wavelengths than that with which the material was illuminated due to thermal (phonon) interactions with the excited atoms.[2] This process was named anti-Stokes fluorescence opposed to Stokes fluorescence in which the emitted wavelength is larger than the absorbed one. Since anti-Stokes fluorescence involves the emission of higher energy photons than those which are absorbed, the net anti-Stokes fluorescence can cause removal of energy from the illuminated material and, as a consequence, its refrigeration. Net radiation cooling by anti-Stokes fluorescence in solid materials was observed experimentally for the first time in 1995 by Epstein's research team in ytterbium-doped fluorozirconate $ZrF_4$-$BaF_2$-$LaF_3$-$AlF_3$-$NaF$-$PbF_2$ (ZBLANP) glass.[3] In this experiment a $Yb^{3+}$-doped sample of ZBLANP in the shape of rectangular parallelepiped of volume 43 $mm^3$, was laser pumped at a wavelength of 1015 nm and cooled to 0.3 K below room temperature with anti-Stokes fluorescence. Since the first experimental demonstration of net cooling of solids in 1995, laser-induced cooling has been observed in a wide variety of glasses and crystals doped with ytterbium ($Yb^{3+}$), thulium ($Tm^{3+}$), and in erbium ($Er^{3+}$) ions.[4] The first observation of cryogenic operation in an all-solid-state refrigerator was reported by Sheik-Bahae's research team at the University of New Mexico in 2010.[5] A temperature drop of ~150 K was demonstrated in a 0.2 $cm^3$ ytterbium doped fluoride crystal ($Yb^{3+}$:$LiYF_4$) at a record cooling power of 110 mW. In all these experiments fluorescence (*incoherent* radiation) is involved in the cooling process and all excited ions radiate independently and do not interact with each other. In 1954, when lasers were not yet discovered, Dicke theoretically predicted the phenomenon of spontaneous *collective* emission of *coherent* photons by an ensemble of excited particles coupled by radiation and noise field.[6] This collective emission is named super-radiance (SR). It was proven theoretically that SR like fluorescence could be used in anti-Stokes regime for laser cooling of solids.[7]

In this paper we consider theoretically laser cooling of an $Yb^{3+}$ doped ZBLAN fiber with SR, which is coherent spontaneous radiation, in comparison with the steady-state cooling with anti-Stokes fluorescence, which is incoherent spontaneous emission. We investigate how coherence, which is not caused by stimulated emission, influences the cooling process. The samples are pumped with a pulsed laser at a wavelength of 1015 *nm*, which is longer than the mean fluorescence wavelength of 999 *nm* of ytterbium ions in a ZBLAN sample. We show that dramatic increase in the rate at which a SR pulse de-excites the upper manifold of the ion not only increases the rate of cooling of a sample but removes a restriction on the use of low phonon energy host towards hosts with higher phonon energies. A theoretical description of the cooling process is presented in Section 2. The results of the simulations are discussed in details in Section 3.

## II. THEORETICAL ANALYSIS

It is well known that incoherent fluorescence in rare-earth (RE) doped glasses or crystals is a result of spontaneous relaxation of *independent* ions in a transparent host. Its intensity, $I$, is proportional to a number of excited RE ions, $N$, in a sample, *i.e.* $I \sim N$. Dicke in his theoretical work considered the entire collection of two-level atoms as a single quantum-mechanical system.[6] He found that under certain conditions, which will be discussed later, the atoms cooperate and relax to the ground state in a time, $\tau_{SR}$, much shorter than the spontaneous relaxation time, $\tau_s$. It has been shown that $\tau_{SR} \approx \tau_s/N$. As a consequence, the radiation intensity is proportional to the square of the number of excited atoms, $I \sim N^2$.[6] The system exhibiting such a cooperative effect was named "super-radiant". A unique property of SR is that all the energy stored in the sample is released in the form of *coherently* emitted light. The source of this coherence is the excited ions correlated over the electromagnetic field. The traditional source of coherency, that is stimulated emission, is *not* involved in this process.

The quantum mechanical theory of SR in RE doped hosts has been developed in Ref.[8] In another paper[7] this theory has been enhanced for a case, when the sample is pumped in the long wavelength tail of the absorption spectrum of RE ions and cooling of the sample can be observed. The experimental history of SR began in 1973, when SR was observed in hydrogen fluoride (HF) gas.[9] The first solid-state experiments on SR were carried out in the early 1980s. $O_2^-$ centers in KCl crystals[10,11] and impurity pyrene molecules in di-phenyl crystals were used in these experiments.[12] In 1999, SR was observed in a RE doped crystal, $Pr^{3+}$:$LaF_3$, for the first time.[13] To lower the SR threshold, this crystal was cooled down to 2.2 K to compensate for the low ions concentration in the sample, as will be shown later.

It is important to emphasise that SR takes place only under certain special conditions imposed on the pump pulse duration, shape and size of the sample. These conditions were presented and discussed in Ref.[6] and is briefly described by the following conditions:

$$\tau < \tau_c < \tau_s, \tau_2, \qquad (1)$$

$$\tau_p < \tau_D, \qquad (2)$$

where $\tau$ is the time of flight of a photon through the sample. $\tau_c$ is the correlation self-formation time in the medium. It characterizes the width of the SR impulse at the half of the peak intensity. $\tau_2$ is the time of phase irreversible relaxation. $\tau_p$ is the duration of the pump pulse. $\tau_D$ is the delay time of SR pulse. The process of SR pulse formation begins as simple incoherent fluorescence when ions in the host do not interact with each other. Gradually interaction of ions through the field causes the correlation of the dipole moments of the ions, which reaches a maximum at $\tau_D$. The left inequality in (1) indicates that propagation times of photons in the sample have to be less than all the characteristic times of the system. The right inequality in (1) indicates that the formation of an SR pulse in the system has to be a faster process than the process of spontaneous relaxation resulting in incoherent emission. It is important to emphasise that the process of formation of an SR pulse is entirely different from the process of formation of the amplified signal in a laser system, which is based on stimulated emission, although in both cases we have coherent radiation. In Ref.[6] it has been shown that in the SR regime a major portion of power is radiated into small angles along the direction where the sample is most extended (Fig.1). A SR signal with a sharp directionality in space has been observed simultaneously in the direction of propagation of the pump signal and in the counter-propagating direction in a RE doped sample.[6] The instantaneous power of the SR signal is given by the equation[8]

$$P(t) = \frac{hc}{\tau_s \lambda_F} \mu \frac{N^2}{4} sehc^2\left(\frac{t - \tau_D}{2\tau_c}\right), \qquad (3)$$

where $t$ is the time, $\lambda_F$ is the mean fluorescence wavelength, $N$ is the number of the excited RE ions, and $\mu$ is the geometric parameter of the medium, investigated in Ref.[14] For a cylindrical geometry of a sample

$$\mu = \frac{\lambda_F^2}{2A_{eff}\left[1 + \sqrt{1 + F^{-2}}\right]}, \qquad (4)$$

where $A_{eff}$ is the effective mode area, $F = A_{eff}/(\lambda_F L)$, $L$ is the length of the sample. The full SR energy radiated by a sample pumped with a single pump pulse can be easily obtained by integrating equation (3), as,

$$E_{SR} = \frac{hc}{\lambda_F} \frac{\tau_c}{\tau_s} \mu \frac{N^2}{2}\left(1 + \tanh\left(\frac{\tau_D}{2\tau_c}\right)\right). \qquad (5)$$

Trivalent ytterbium ions can be described with a two-level model for absorption and emission processes between the ground-state, $^2F_{7/2}$, and the excited-state, $^2F_{5/2}$ manifolds. As it is shown in Ref.[6] the correlation self-formation time can be calculated as $\tau_c = \tau_s/(N\mu)$, and the delay time of a SR pulse can be obtained as $\tau_D = \tau_c \ln(N\mu)$. The population density in the excited state manifold, $N_2$, changes with time and has to satisfy the relation

$$\frac{dN_2}{dt} = \frac{P_p(t)\lambda_p}{A_{eff}hc}[\sigma_{abs}(\lambda_p)N_1(t) - \sigma_{se}(\lambda_p)N_2(t)] - \frac{N_2(t)}{\tau_s}, \qquad (6)$$

where $P_p(t)$ is the instantaneous pump power at a wavelength $\lambda_p$ at the instance $t$, $\sigma_{abs}(\lambda_p)$ and $\sigma_{se}(\lambda_p)$ are absorption and emission cross-sections at the pump wavelength, $\lambda_p$. $N_1(t)$ is the population density of the ground-state manifold at the time, $t$. $N_1(t) + N_2(t) = N_T$ is the total number density of Yb$^{3+}$ ions. Using equation (6) it is easy to calculate that if a rectangular pulse with power $P_p^{amp}$ and of duration $\tau_p$ is applied to the sample the number of excited ions in the sample is

$$N = L\sigma_a N_T \tau_p P_p^{amp} \frac{\lambda_p}{hc}, \qquad (7)$$

where $L$ is the length of the sample. As a result of thermalization some electrons in the excited state manifold move from the bottom to the top of the manifold absorbing the energy of the phonons responsible for heat generation in the sample (insert in Fig. 2). Anti-Stokes emission of photons follows with mean photon energy $h\nu_F$, where $\nu_F = c/\lambda_F$ is the mean frequency of anti-Stokes photons, which is higher than that of the absorbed photon. This anti-Stokes emission removes energy from the sample providing its cooling. In the case of cooling with incoherent fluorescence the cooling energy removal is in the time of order ~ $\tau_s$, the time of spontaneous relaxation, *i.e.* the power of the anti-Stokes fluorescence is

$$P_{SP} = Nhc/(\tau_s\lambda_F). \qquad (8)$$

This is the energy removed from the sample every second with anti-Stokes fluorescence.
As one can see from Eq. (5) the average power of the SR pulse emission is

$$P_{SR} = \frac{hc}{\lambda_F}\frac{\tau_c}{\tau_s}\frac{1}{\tau_D}\mu\frac{N^2}{2}\left(1+\tanh\left(\frac{\tau_D}{2\tau_c}\right)\right). \qquad (9)$$

This is the energy removed from the sample every second with SR. Comparing equations (8) and (9) one can introduce an effective radiative relaxation time, which characterizes the rate of electron relaxation in the case of SR emission: $\tilde{\tau}_s = \tau_s/C$, where

$$C = \frac{\tau_c}{\tau_s}\mu\frac{N}{2}\left(1+\tanh\left(\frac{\tau_D}{2\tau_c}\right)\right). \qquad (10)$$

In this case $P_{SR} = CP_{SP}$. If $C > 1$, the energy leaves the sample significantly faster with SR than with fluorescence used in traditional cooling experiments. Indeed, every second, SR removes energy from the sample, which is a factor $C$ greater than the energy removed with fluorescence. The efficiency of cooling with SR can be determine as the ratio of the energy removed from the sample with a SR pulse to the energy absorbed by the sample from the pump pulse. The energy absorbed by the sample from the pump pulse is $Nh\nu_p$. Using (5) one can calculate the efficiency of cooling with SR as

$$\eta_{cool}^{SR} = \frac{\nu_F}{\nu_p} \frac{\tau_D}{\tau_s} C - 1. \tag{11}$$

Using (10) one can conclude that the efficiency of cooling with SR is approximately equal to the efficiency of cooling with the anti-Stokes fluorescence, that is $\eta_{cool}^{SP} = \nu_F/\nu_p - 1$, which is calculated in Ref.[15, 16], if non-radiative relaxation in the host is not taken into account. It is important to emphasise that for a SR pulse the number of excited ions in the sample has to be in excess of a threshold value, which can be calculated with the right inequality in (1) taking into account that $\tau_c = \tau_s/(N\mu)$:

$$N_{th} > \frac{\tau_s}{\tau_2 \mu}. \tag{12}$$

This threshold value can be reduced by increasing the pump power, $P_p^{amp}$, or total ion density $N_T$. The threshold value, $N_{th}$, will decrease with the temperature of the sample, since $\tau_2$ increases as the temperature drops.

## III. RESULTS AND DISCUSSION

In this part of the paper computer simulations describing the performance of the proposed scheme are presented. Since SR can be observed in samples highly *extended* in one direction, we consider in our simulations a cylindrical ZBLAN sample doped with $Yb^{3+}$ ions with a radius, $r = 0.5$ *mm*, and a length $L = 5$ *mm*. The refractive index of the sample is $n = 1.5$. The spectra of absorption and emission of $Yb^{3+}$ ion in ZBLAN have been presented in Ref.[17], and illustrated in Fig. 2. The sample is pumped with a pulsed laser at the wavelength $\lambda_p = 1015$ *nm*, which is longer than the mean fluorescence wavelength of the $Yb^{3+}$ ions in ZBLAN host, ($\lambda_F = 999$ *nm*). SR is a cooperative process and requires the ion concentration to be high. We consider a sample with $N_T = 1.45 \times 10^9$ *ions/µm³*. As one can see from Eq. (7) that if the pump power $P_p \approx 433.5$W and the duration of the pump pulse $\tau_p = 10$ *ns*, the number of the ions in the sample participating in the cooling process with SR is approximately equal to $N = 6 \times 10^{12}$. This value satisfies (12).

     We want to show that the coherence introduced in the cooling process with an SR pulse permits a dramatic increase in the rate of cooling increasing the efficiency of the process in comparison with traditional cooling with anti-Stokes fluorescence of low phonon hosts. But first of all we must prove that the inequalities (1) and (2) are satisfied for the sample under consideration and an SR pulse can be generated in the system. Equation (3) describes the time evolution of the power of the SR. The SR power, normalized to its maximum value, as a function of time is illustrated in Fig. 3. It has a maximum, which corresponds to the delay time of SR pulse, $\tau_D \approx 14$ *ns*. The delay time $\tau_D$ satisfies the relation (2). The correlation self-formation time in the medium characterizing the SR impulse at the half of peak intensity is $\tau_c \approx 1$ *ns*. This satisfies the relation (1). Indeed the time of energy relaxation is $\tau_s \approx 1.9$ *ms* and the time of flight of a photon through the sample is $\tau = Ln/c \approx 25$ *ps*. It is important to emphasise that as soon as the left inequality in the relation (1) is satisfied, the system is free from re-absorption, since all generated photons leave the sample. The delay time of the SR pulse and the correlation self-formation time change with the value of the pump power, since the number of excited ions, $N$, participating in formation of the SR pulse depends on the value of the pump power. The dependence of the delay time of the SR pulse, $\tau_D$, on the value of power of the pump signal is

illustrated in Fig. 4. The delay time decreases as the pump power increases. The correlation self-formation time, $\tau_c$, is illustrated in Fig. 5. As in the case of the delay time of the SR pulse, the correlation self-formation time decreases with increasing power of the pump pulse.

The parameter $C$ described with Eq. (10) characterises the increase in the rate of cooling with SR in comparison with the rate of cooling with fluorescence. In the case of our sample if the $P_p \approx 433.5W$ the value of $C = 1.4 \times 10^5$. The dependence of $C$ on the pump power is illustrated in Fig. 6. We see that cooling with SR permits a significant increase (by a factor of $\sim 10^5$) in the rate at which energy for cooling leaves the sample, compared to traditional cooling with anti-Stokes fluorescence. An important advantage of cooling with SR is that the cycling time of the cooling process can be reduced considerably. In one or several cooling cycles cryogenioc temperatures can be achieved.

The excited state of an ion can decay non-radiatively via interactions with optical phonons of the host material. This undesirable effect causing heating of the sample can be characterised by the internal quantum efficiency

$$\eta = W_r/(W_r + W_{mp}), \qquad (13)$$

resulting from the competition between radiative, $W_r$, and multi-photon relaxation, $W_{mp}$, rates, where

$$W_{mp} = \frac{W_0}{\left(1 - e^{-\frac{E_p}{\kappa_B T}}\right)^{n_p}}. \qquad (14)$$

$$W_0 = Be^{-\alpha \Delta E} \qquad (15)$$

is the spontaneous transition rate at $T = 0K$ due to the zero-point fluctuations of the phonon field. $E_p$ is the dominating phonon energy. $\Delta E$ is the energy gap that is bridged by the emission of $n_p$ phonons. $B$ and $a$ are material-dependant parameters that have to be determined experimentally. The internal quantum efficiency reduces the cooling efficiency to $\eta_{cool} = \eta \lambda_p/\lambda_F - 1$. Since, in the case of SR, the radiative relaxation time decreases dramatically as $\tilde{\tau}_s = \tau_s/C$, the radiative relaxation rate, $W_r$, increases correspondingly. As it was already mentioned in the case of our sample it increases approximately by a factor of $10^5$. As consequence laser cooling with SR can increase the internal quantum efficiency described with Eq. (13) dramatically in comparison with the case of traditional cooling with anti-Stokes fluorescence for *any host* material. It is important to emphasise that $W_r$, only slightly depends of the host material in comparison with $W_{mp}$. This means that if, in the case of cooling with SR, we save the internal quantum efficiency equal to one in the case of traditional cooling with anti-Stokes fluorescence, the value of $W_{mp}$ can be increased considerably as one can see in Eq. (13). That is, in the case of cooling with SR, we can use *new* hosts with considerably higher maximum phonon energy than traditional hosts used for cooling with anti-Stokes fluorescence saving the value of the internal quantum efficiency, $\eta$. For example, one can cool down a RE-doped silica sample with maximum phonon energy ~1100 cm$^1$ or other hosts with comparable maximum phonon energy with SR saving the value of the internal quantum efficiency comparable with one in the case of cooling of RE-doped low phonon hosts with anti-Stokes fluorescence.

## IV. CONCLUSION

In summary, a radically new approach to laser cooling of solids with SR is presented and comprehensively investigated in comparison with the traditional cooling technique based on anti-Stokes fluorescence. SR is a collective coherent emission of an ensemble of excited ions. In this process independent ions cooperate and relax to the ground state in a time much shorter than the spontaneous relaxation rate. This "fast" relaxation rate is the *crucial point* of the new approach to the cooling process. It permits a dramatic increase in the rate of cooling in comparison with the traditional cooling cycle with incoherent anti-Stokes spontaneous radiation and to reach cryogenic temperatures in a short time increasing the efficiency of the process for any host. It makes possible the use of new materials with higher maximum phonon energy (for example silica), so far considered unsuitable as RE-doped hosts for laser cooling. This scheme can break the limitation on the minimum temperature achievable in RE-doped solids. It is important to emphasise that for the realization of SR some restrictions must be imposed on the shape of a sample. The sample must be greatly extended in one dimension to provide formation of the SR pulse. The time of flight of a photon through the sample must be short in comparison with all other characteristic times of the system. The duration of the pump pulse must be shorter than the delay time of SR pulse.

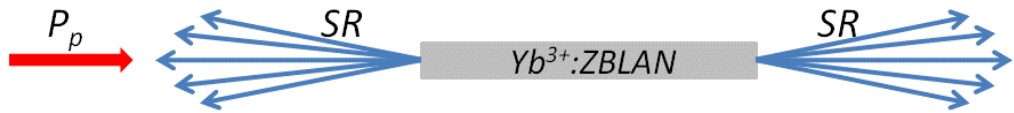

Fig. 1 Structure under consideration.

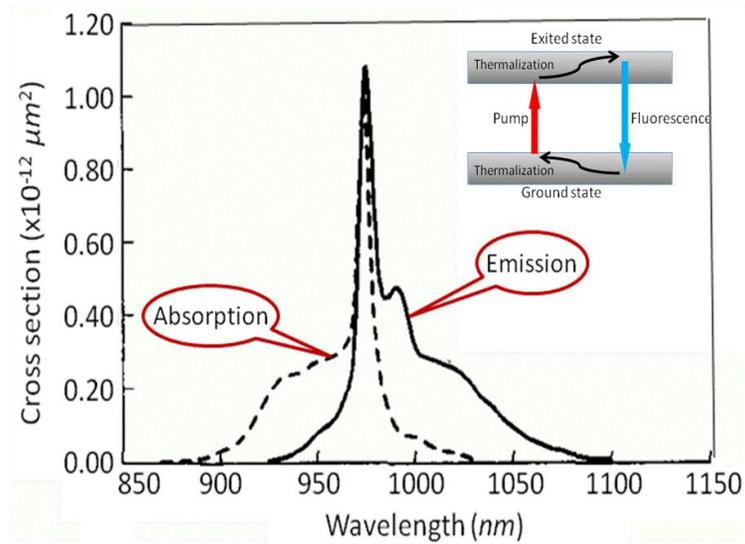

Fig. 2. Absorption and emission cross-sections at room temperature of Yb$^{3+}$:ZBLAN. The insert illustrates an energy diagram showing the routes to anti-Stokes cooling.

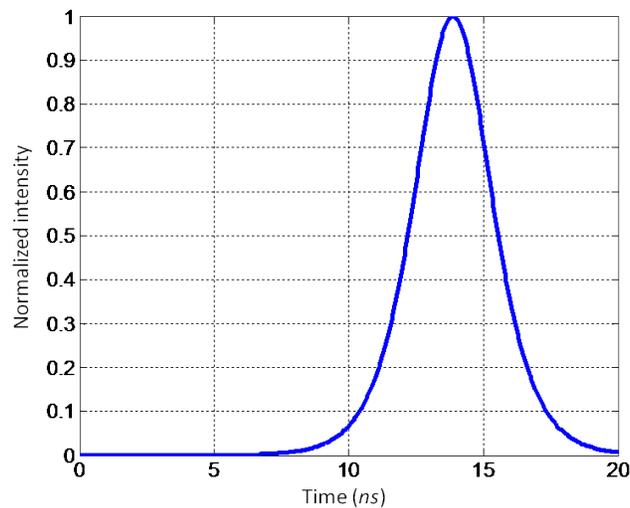

Fig. 3. Normalised SR intensity as a function of time.

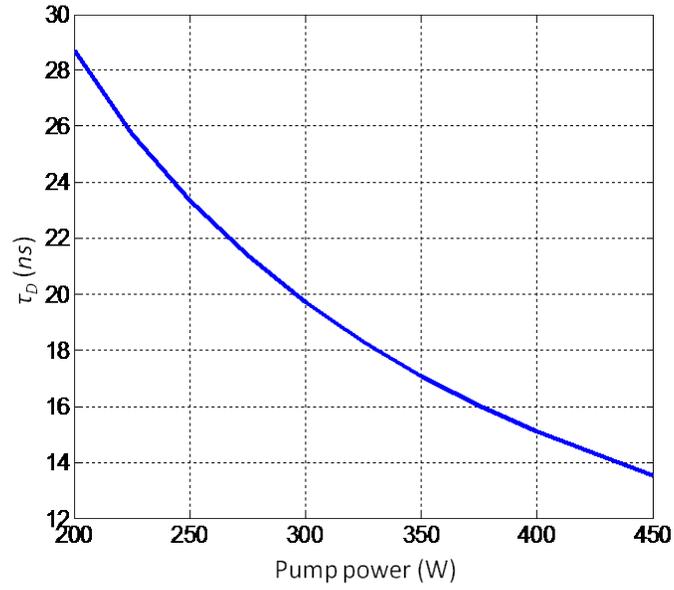

Fig. 4. Dependence between the pump power and the delay time of SR pulse, $\tau_D$. $\tau_p = 10$ ns.

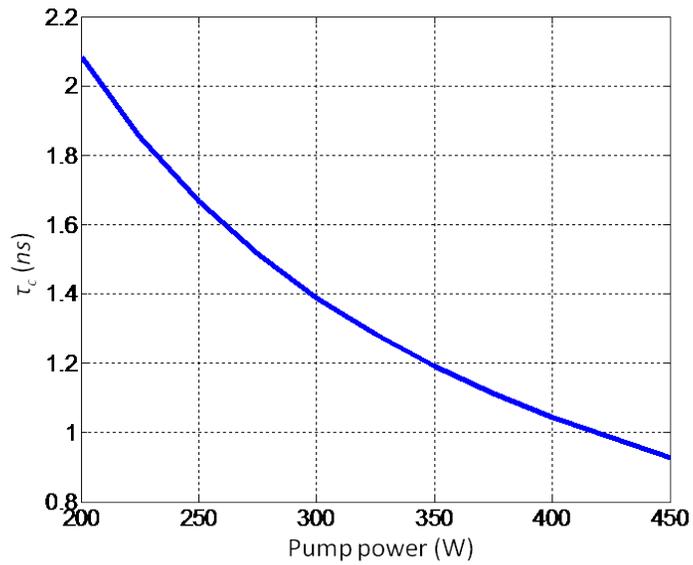

Fig. 5. Dependence between the pump power and the time of correlation self-formation, $\tau_c$. The pump pulse duration, $\tau_p = 10$ ns.

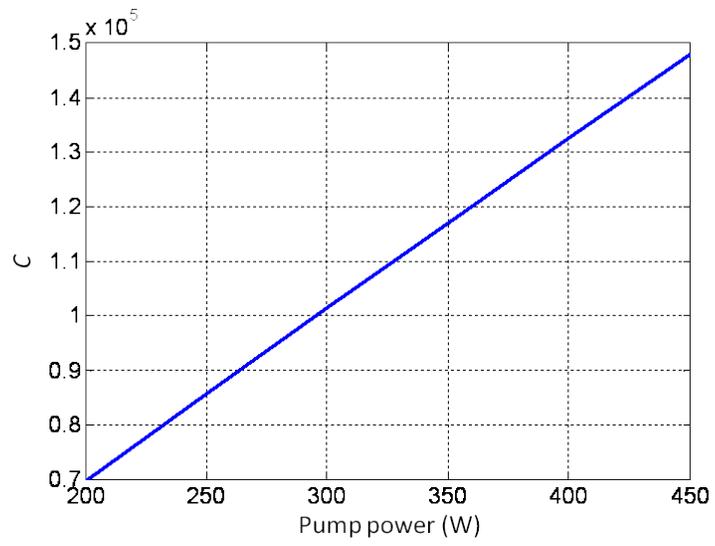

Fig. 6. Dependence between the pump power and the coefficient "$C$", with $\tau_p =10$ *ns*.